\documentclass[prb,twocolumn,floatfix,showpacs]{revtex4}
\usepackage{graphicx}
\begin{document}

\title{Spin Stiffness of  Stacked Triangular Antiferromagnets}

\author{A. Peles and B.W. Southern}
\email[email:]{souther@cc.umanitoba.ca}
\affiliation{Department of Physics and Astronomy\\ University of Manitoba \\ Winnipeg Manitoba \\ Canada R3T 2N2}

\date{\today}

\begin{abstract}
We study the spin stiffness of stacked triangular antiferromagnets using both  heat bath and broad
histogram Monte Carlo methods. Our results are consistent with a continuous transition belonging
to the chiral universality class first proposed by Kawamura.
\end{abstract}


\maketitle

\section{Introduction}

The magnetic ordering of geometrically frustrated antiferromagnets differs 
substantially  from the conventional magnetic ordering in  nonfrustrated magnets\cite{n1,n2}.
Indeed, the nature of the phase transition in the case of stacked triangular antiferromagnets
remains controversial\cite{n3,n4,n5,n6,n9,n10,n11,n17,n18,n19}. 
The geometry of the stacked triangular lattice has  triangles as the elementary units
and this arrangement inhibits an anti-parallel alignment of the spins in each triangular layer.
Consequently, the system is said to be geometrically frustrated.
This frustration leads to a comprise where the spins on each triangle adopt a   non-collinear spin ordering at
low temperatures. The spins form a planar configuration in which nearest neighbours
are oriented at an angle of 120 degrees with respect to one  another. 
The ground state can be described by a matrix like order parameter giving the 
orientation of each spin on the elementary triangles and forms an SO(3) parameter space\cite{n7}. 
This unusual symmetry of the order parameter and the appearance of
'chiral' degrees of freedom  which correspond to the ground state having two possible
realizations, left and right handed, lead Kawamura\cite{n8} to
conjecture the existence of a new chiral universality class . The chiral degrees of freedom are believed to be
responsible for the novel critical behaviour but they are not decoupled from the spin degrees of freedom and
the two quantities order simultaneously.
While recent field-theoretic renormalization group studies of this system  using 
an expansion up to six loops in fixed dimension $d=3$ 
indicate the existence of a stable fixed point that corresponds  to the proposed chiral
universality class\cite{n17,n18},  similar analyses
 using  a three loop perturbation technique as well as an epsilon
expansion approach to the  same order, did not find a stable fixed point and hence
exclude the possibility of  a continuous phase transition for 
this frustrated system\cite{n9,n10}. 
Non perturbative RG approaches find that the phase transition is possibly a very weak 
first order  transition
with  effective critical exponents\cite{n11}. 

In the present work we use both a standard heat bath Monte Carlo method as well as a recently
developed broad histogram method\cite{n14} to study the classical isotropic antiferromagnet on this geometry.
In particular, we study the spin stiffness which provides a direct measure of the correlation length
exponent $\nu$. The spin stiffness is a convenient quantity to study since it does not require knowledge of
the order parameter but does measure the rigidity of the ordered phase against fluctuations. 
Our results confirm the picture of a continuous transition belonging to a new chiral
universality class.

\section{Model and Methods}

The model is described by the following Hamiltonian
\begin{equation}
H = -\sum_{i<j} J \vec{S}_{i} \cdot \vec{S}_{j} -
\sum_{k<l} J' \vec{S}_{k} \cdot \vec{S}_{l}.
\end{equation}
where $\vec S_{i}$ is a classical three component vector of unit length 
located at the sites $i$ of a hexagonal lattice. 
The first sum is over 
nearest neighbours in the triangular planes which interact with an
antiferromagnetic coupling $J <0$
and the second sum is over inter-plane nearest neighbours which are taken to have a
ferromagnetic coupling $J' >0$ with $|J'| =|J|=1$.   Hence all energies and temperatures are measured in units of $|J|$.

We study the response of the system to a virtual  twist of the spin system. The spin  stiffness , or helicity modulus\cite{n13} ,  measures the increase in  free energy associated with twisting the
order parameter   in spin space  by imposing  a gradient of the twist angle about some axis $\hat{n}$ in spin space
along some  direction $\hat{u}$ in the
lattice. The spin  stiffness   can be written 
 as a second derivative  of the free energy
 with respect to the strength of the gradient  and can be calculated as an equilibrium response function\cite{n12} . Finite size scaling theory predicts that the spin stiffness should vanish at the critical point with an exponent related to the correlation length exponent.

We  calculate the diagonal elements of the spin  stiffness tensor
corresponding to twists about three principal directions in spin space.  If we divide the lattice sites into three
equivalent sublattices $A, B$ and
$C$ corresponding to the vertices of the elementary triangles, then a chirality vector can be defined 
to characterize the  non-collinear ordering of the spins. The chirality is defined locally for each upward
(downward) triangle by the following expression
\begin{equation}
\vec{K}_{\triangle}=\vec{S}_{A}\times\vec{S}_{B}+\vec{S}_{B}\times\vec{S}_{C}+
\vec{S}_{C}\times\vec{S}_{A}
\end{equation}

In the ground state, the chirality is uniform and perpendicular to the spin planes.
This symmetry of the order parameter suggests that the
average chirality  direction ($\hat{K}$) be chosen for one of the principal axes and the other two directions
($\hat{\perp}_{1},\hat{\perp}_{2}$) are chosen to be
in the spin plane perpendicular to the average chirality vector such that the three vectors form an orthonormal triad. 
The spin stiffness component $\rho_{\alpha}$ at temperature $T$ can be written as \cite{n12}

\begin{widetext}
\begin{equation}
\rho_{\alpha}=
\frac{1}{N}\sum_{i<j} J_{ij} (\hat{e}_{ij} \cdot \hat{u})^{2}
\left < S_{i}^{\beta}S_{j}^{\beta} + 
S_{i}^{\gamma}S_{j}^{\gamma} \right>
- \frac{1}{NT} 
\left < \left ( \sum_{i<j} J _{ij} (\hat{e}_{ij} \cdot \hat{u})
 \left[
S_{i}^{\beta}S_{j}^{\gamma} - S_{i}^{\gamma}S_{j}^{\beta}\right]
\right)^{2}\right>
\end{equation}
\end{widetext}
where $\alpha,\beta,\gamma= \hat{K},\hat{\perp}_{1},\hat{\perp}_{2}$ and the indices are taken in cyclic order. 
 The twist is taken to be along the   $\hat{u}$ direction in the lattice   and $\hat{e}_{ij}$ is a unit vector directed along the nearest neighbour bond from site $i$ to $j$ . The angular brackets indicate a thermal average in the canonical ensemble. Since the ground state is a planar spin arrangement, the stiffnesses satisfy a perpendicular axis theorem
$\rho_{\hat{K}}= \rho_{\hat{\perp}_{1}} + \rho_{\hat{\perp}_{2}}$ at zero temperature. Deviations from this relationship are
a measure of fluctuations of spins from the planar order.

We perform numerical simulations using both a conventional Monte Carlo (MC)
heat bath method and the more recent broad histogram method (BHM) introduced by Oliveira et. al.
The latter method is based on the microcanonical ensemble approach to statistical sampling
 at fixed energy and allows an accurate estimate of the energy density 
of states\cite{n14,n15,n16} $g(E)$ .
By knowing the density of states $g(E)$ and the microcanonical averages of various
 quantities $<Q>_E$, their  temperature dependence  can be 
determined 
by using the following expression for the canonical averages 
\begin{equation}
<Q>_{T}=\frac{ \sum_{E}<Q>_{E}g(E)e^{-E/T}}{\sum_{E}g(E)e^{-E/T}}
\end{equation}

In the conventional heat bath method temperature is tuned as an  external 
parameter 
and number of  temperature points is limited by number of computer runs. The BHM method allows us to probe the
system in a continuous range of $T$ but requires a large number of energy bins for large system sizes.
We simulate spin systems of size $N=L^3$ with $L=24, 30, 42,60$ 
and $66$ for the heat bath method and only up to $L=60$ for the BHM method. Periodic boundary conditions
are employed for both methods.
 We find excellent agreement between these two numerical methods.

\section{Results}

\begin{figure}
\centering
\includegraphics[width=3.5in,angle=0]{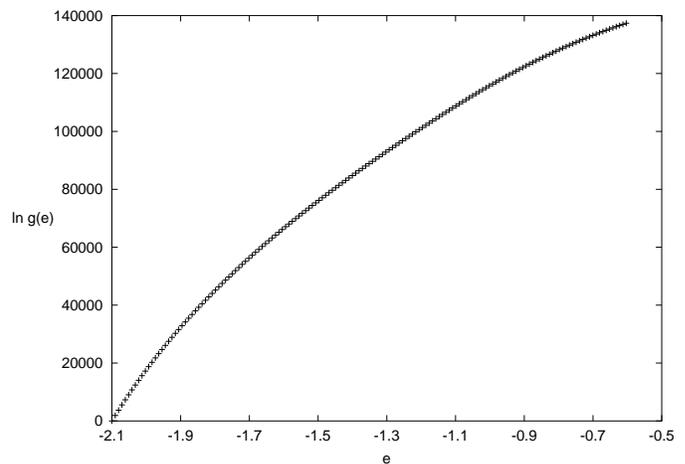}
\caption{Natural logarithm of the energy density of states versus the energy per site $e$ for a $42 \times 42 \times 42$ lattice in
the range $-2.1 \le e \le -0.5$.}
\end{figure}

\begin{figure}
\centering
\includegraphics[height=2.5in]{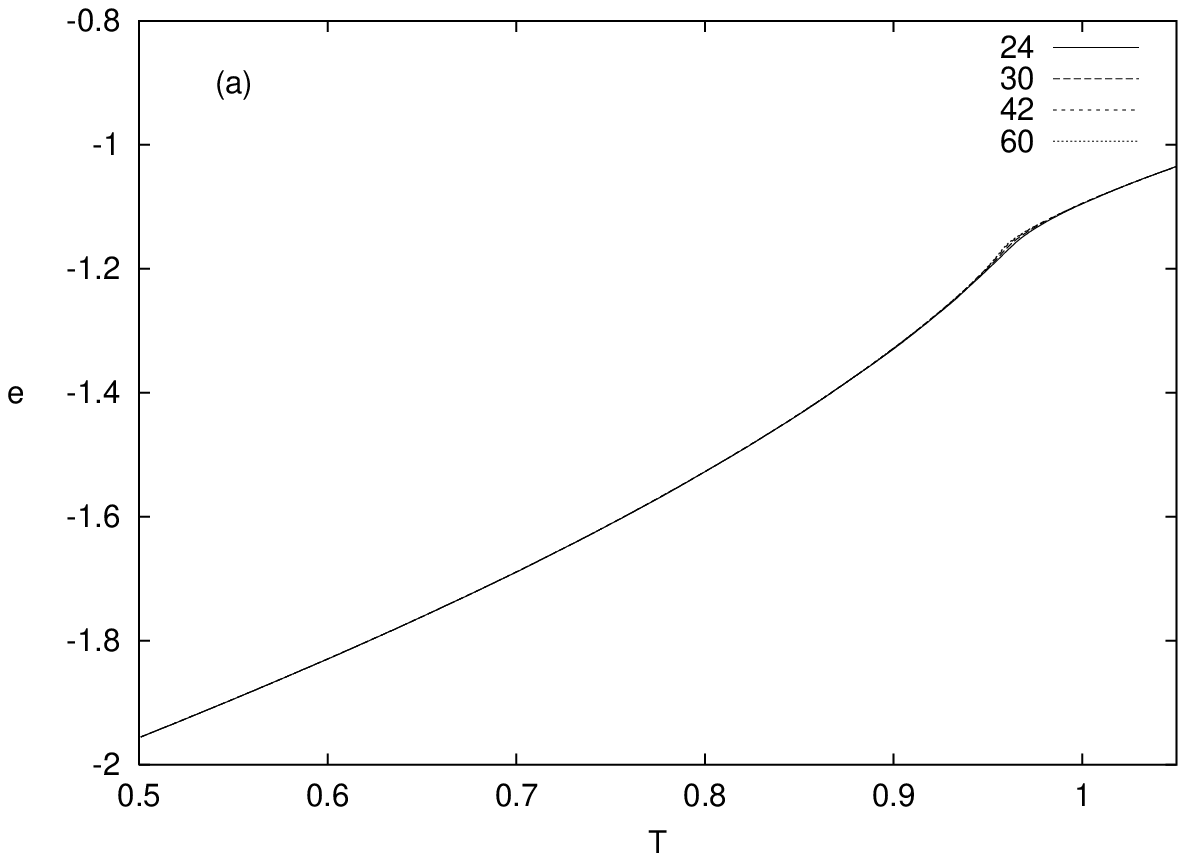}
\includegraphics[height=2.5in]{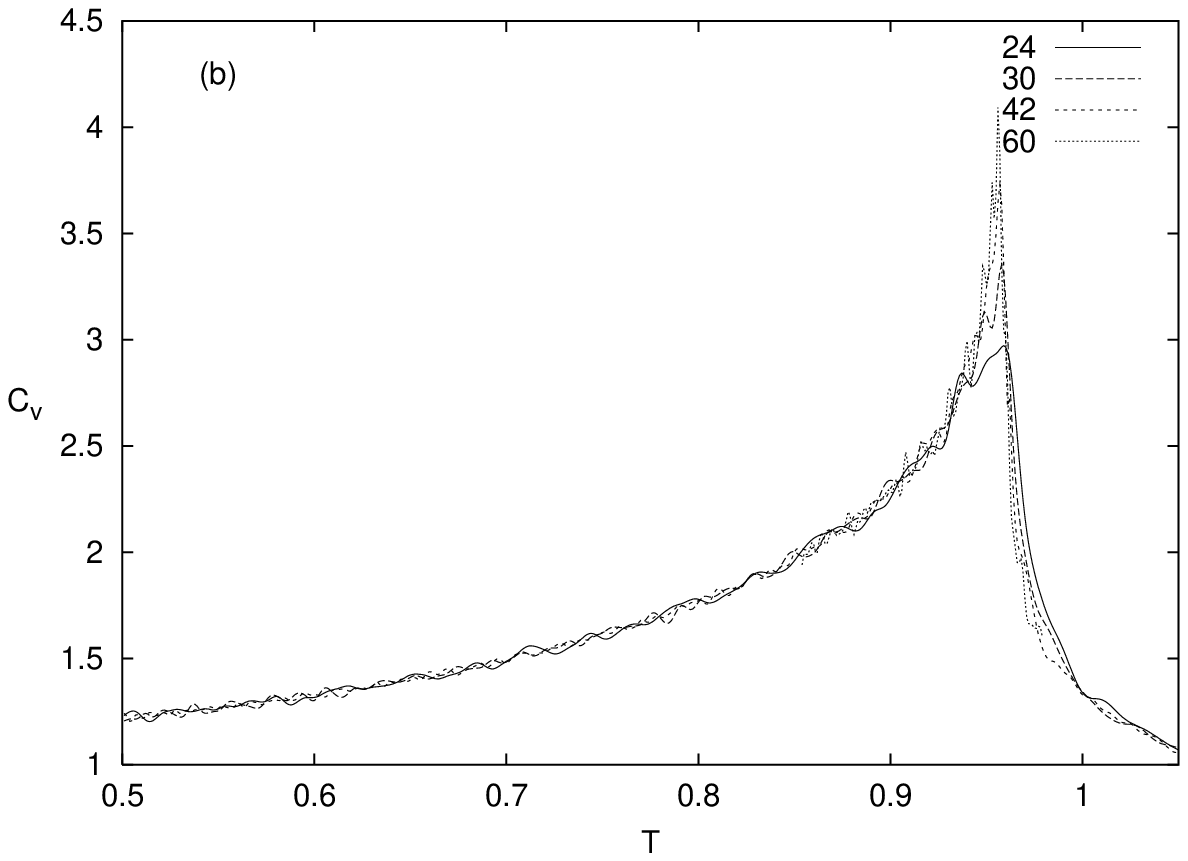}
\caption{Energy per site $e$ and specific heat  $C_v$ obtained using the BHM method for $L=24,30,42,60$.}
\end{figure}

The broad histogram method(BHM) is based on the relation
\begin{eqnarray}
g(E) < N_{up}(E) > = g(E+ \Delta E) < N_{dn}(E+ \Delta E) >
\end{eqnarray}
where $<N_{up}(E)>, <N_{dn}(E)>$ are microcanonical averages which measure the number of moves which increase (decrease)
the energy by the amount $\Delta E$. Once these microcanonical averages are known, the microcanonical temperature $T_m(E)$
can be determined from
\begin{eqnarray}
1/T_m(E) & \equiv &  \frac{d \ln g(E)}{dE} \nonumber \\
&\simeq & \frac{1}{\Delta E} \ln \frac{<N_{up}(E)>}{<N_{dn}(E+ \Delta E)>}
\end{eqnarray}
and we can then integrate this expression in some range of energy to obtain the energy density of states $\ln g(E)$ as a function of $E$.
In our case the energy is a continuous variable and we divide the energy axis into  bins of a fixed size $\Delta E = 1.8 $
such that  $\Delta E << E$, where $E$ is {\it total} energy of interest.
We employ  a simple microcanonical dynamics to sample phase space and  the energy 
density of states $g(E)$ (up to a multiplicative constant) is determined using the  BHM relation above. 
One microcanonical sweep consists of a random sweep through the lattice sites and generating a
new configuration of the spins by restricting the choice of a new random orientation of the spin at site $i$ with
respect to the local field of the nearest neighbours such that the total energy of the system remains
within the energy interval  $E, E + \Delta E$. 
At any given value of $E$ , 75 microcanonical sweeps 
were performed and 25 sample measurements were taken of various thermodynamic quantities such as the energy, specific heat and spin stiffness.
Before sampling the next energy interval, 40 initial microcanonical sweeps were performed to avoid
correlations. This procedure was repeated using different seeds for random numbers
and errors were determined using the standard deviations for these separate measurements.

Figure 1 shows our results for  $\ln g(e)$ as a function of $e=E/N$ in the case of a $42 \times 42 \times 42$ lattice. The units are
arbitrary since we integrate equation (6) starting from $e = -2.1$  and not the ground state value $e_0 =-2.5$ .The number of
energy bins used for this energy range was 61740. For general values of $L$, the number of energy bins required to study this same range with the same fixed size of energy bin is $5 L^3 /6$ and is thus of the same order as the number of sites.
When the energy density of states is combined with the microcanonical averages $<Q>_E$ for various thermodynamic quantities, we can then plot
them as continuous functions of $T$ using equation (4). Figure 2 shows the energy per site and the specific heat obtained using the BHM
method for various linear sizes $L$.  The energy displays strong finite size effects near the temperature where the specific heat has a maximum. The figures clearly indicate that a phase
transition occurs near $T  \sim 0.96$ in agreement with previous MC studies.\cite{n1}

\begin{figure}
\centering
\includegraphics[height=2.5in,angle=0]{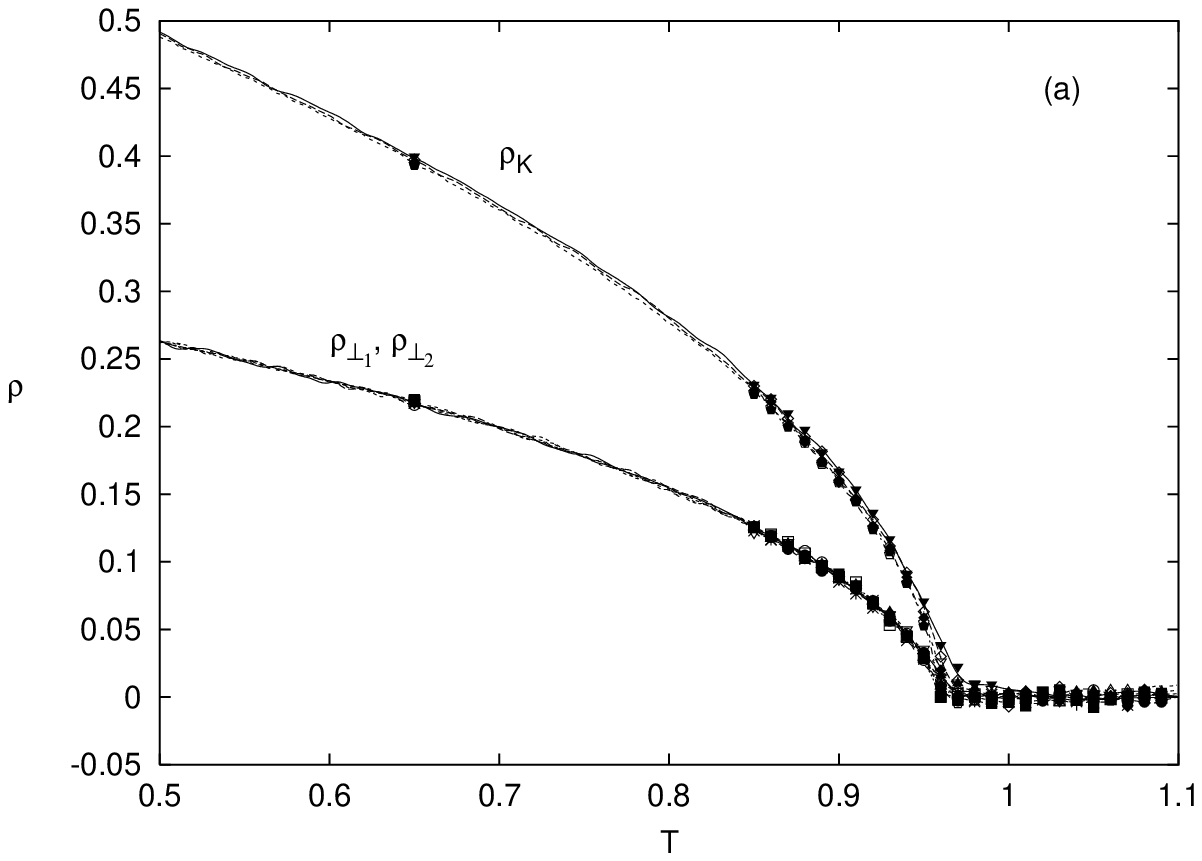}
\includegraphics[height=2.5in,angle=0]{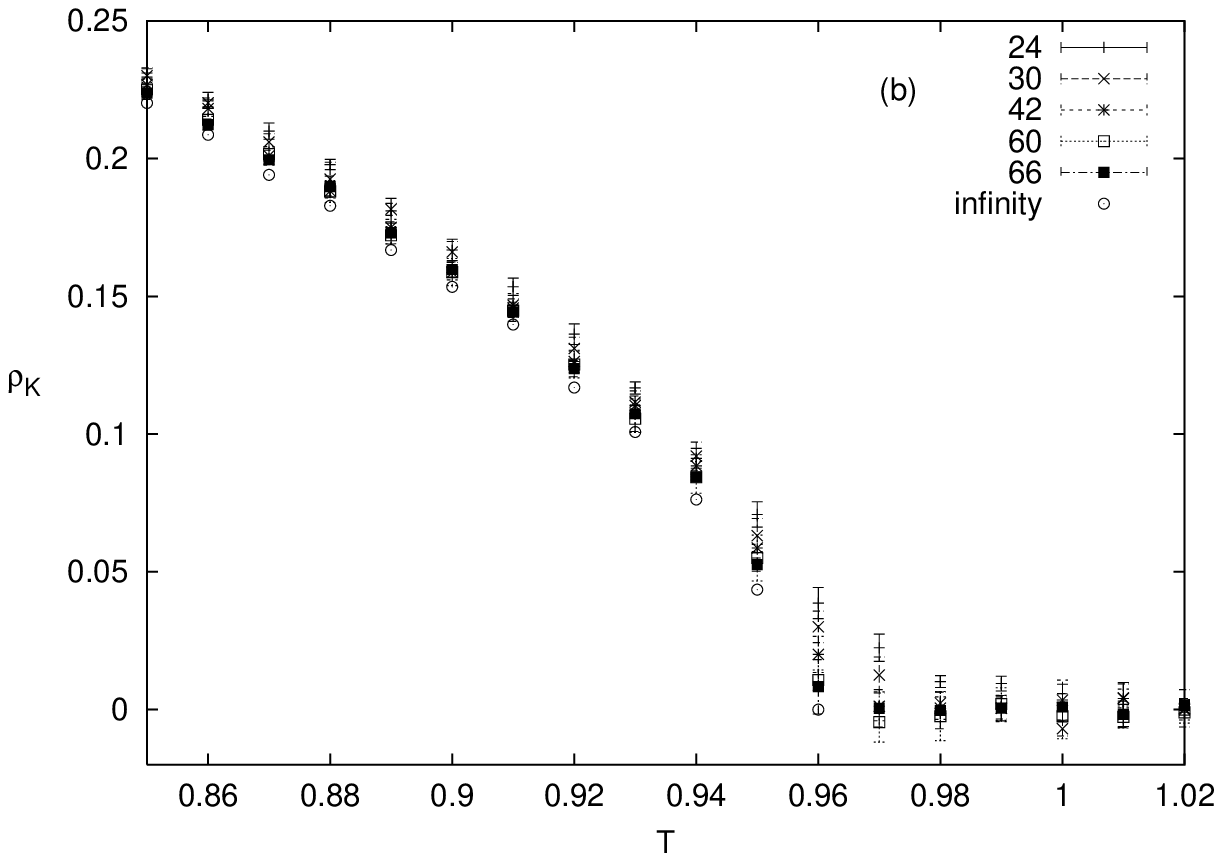}
\caption{Spin stiffnesses as a function of $T$.  a) the points indicate the heat bath results and the lines correspond to the BHM results. All three stiffnesses vanish at the same finite temperature near $T \sim .96$.  b) the heat bath results for $\rho_K$ in a smaller temperature range show significant finite size effects near $T_c$.}
\end{figure}

\begin{figure}
\centering
\includegraphics[width=3.5in]{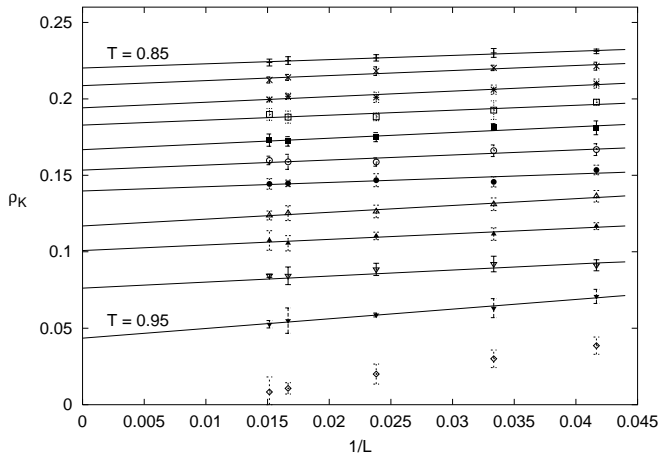}
\caption{The same data as in figure 3(b) is plotted as a function of $1/L$ for a set of equally spaced temperatures in the range
$ .85 \leq T \leq .95$. 
Extrapolation to the large $L$ limit yields estimates for $\rho_K$ for an infinite lattice.}
\end{figure}

We have used both the BHM method and a Monte Carlo heat bath method at fixed values of $T$
to calculate the spin stiffness. In the heat bath method, we discard the first 1000  sweeps 
and perform 45000  MC steps in each  run.
Figure 3(a) shows both our heat bath results, indicated by points, and the BHM results, indicated by lines, for the three
stiffnesses for various lattice sizes $L$ as a function of the temperature $T$.  The relation $\rho_{\hat{K}}= \rho_{\hat{\perp}_{1}} + \rho_{\hat{\perp}_{2}}$ is well satisfied for all values of $T < 0.95$ indicating that there is a relatively small deviation
from the planar spin configuration. All three stiffnesses are nonzero
at low $T$ and vanish near $T \sim .96$ which corresponds to the specific heat divergence in figure 2.
 Figure 3(b) shows the  heat bath  data for 
$\rho_{\hat{K}}$ on an enlarged temperature scale. The stiffnesses clearly show large finite size effects and approach zero near 
$T \sim .96$. The points labelled infinity are obtained by plotting $\rho_{\hat{K}}$  versus $1/L$ at various values of $T$
and extrapolating to the large $L$ limit as shown in figure 4. 

\begin{figure}
\centering
\includegraphics[width=3.5in]{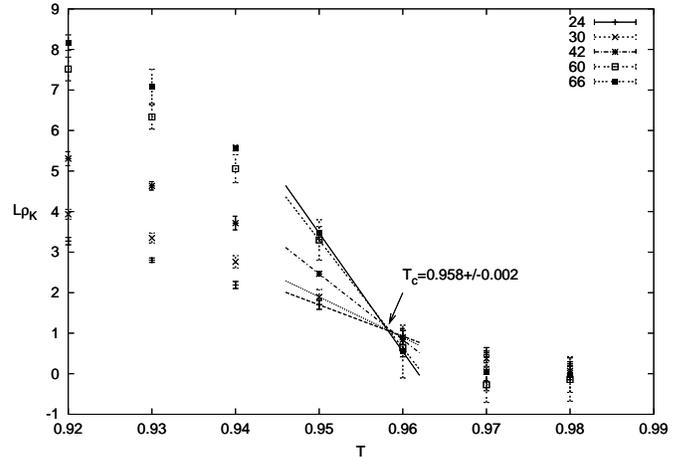}
\caption{$L \rho_K $ versus $T$ for various lattice sizes are indicated by the points. The lines are linear interpolations
which indicate a unique crossing point at $T_c = 0.958 \pm 0.002$}
\end{figure}

These finite size effects can be used to determine the correlation length exponent $\nu$ directly.
Finite size scaling considerations for $\rho(T,L)$ predict 
\begin{eqnarray}
\rho(T,L) = \frac{1}{L} f(L/ \xi) = \frac{1}{L} f(L^{1/ \nu} |t|)
\end{eqnarray}
where $t$ is the reduced temperature.  This form suggests that we can plot $L \rho(T,L)$ versus $T$ to identify $T_c$  as
the temperature where the curves for different values of $L$ intersect. 
Figure 5 shows our heat bath results for $L \rho_K$
as a function of $T$ for lattice sizes $L=24,30,42,60,66$. Linear interpolations of neigbouring temperature points indicate that the curves intersect at a value
of $T_c = 0.958 \pm 0.002$. We have also used our BHM results in the same temperature range and we obtain
the same estimate for $T_c$.

In the limit as $L \rightarrow \infty$,  the scaling form predicts $\rho \sim |t|^{\nu}$. Using the values of the
stiffness obtained by extrapolating to large values of $L$ as in figure 4 and then  plotting these versus
$ |t|$ on a ln-ln scale, we can obtain an estimate of  $\nu$.  Figure 6 shows our results for $\rho_K$ which yields the value
$\nu = .589 \pm .007$. This value agrees very well with previous Monte Carlo estimates\cite{n1} but is slightly larger than the
value found by the recent six loop renormalization
group calculations in three dimensions.\cite{n17,n18}

\begin{figure}
\centering
\includegraphics[width=3.5in]{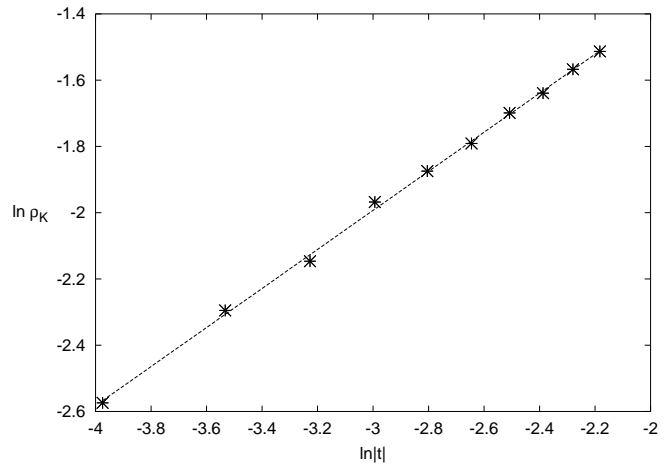}
\caption{A ln-ln plot of $\rho_K$  as $L \rightarrow \infty$ versus $|t|$ using the estimated value of $T_c$ yields a value of $\nu =.589 \pm .007$.}
\end{figure}

\begin{figure}
\centering
\includegraphics[width=3.5in]{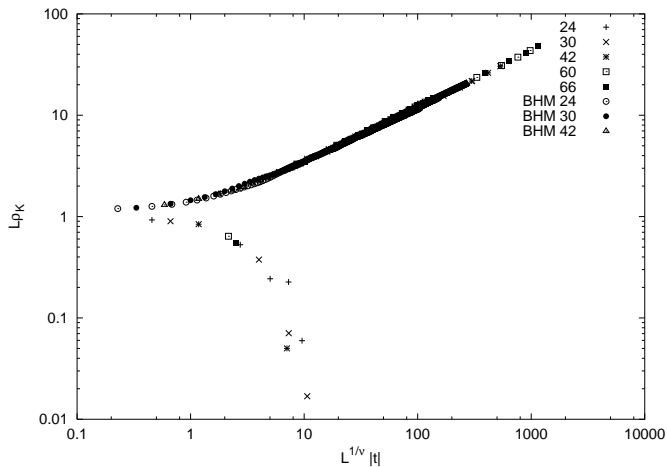}
\caption{Finite-size scaling plot of $L \rho_K$ versus $L^{1/\nu} |t|$ produces a universal curve}
\end{figure}

Figure 7 shows a finite size scaling plot of our stiffness results using the values of $T_c$ and $\nu$ quoted above. The data
obtained from both the heat bath MC method for sizes $L=24,30,42,60,66$ and the BHM method for $L=24,30,42$ collapse very well to
a universal function for temperatures below $T_c$. The value $\nu = .589 \pm .007$ is certainly very different from the value 
$\nu= 0.7113$ which describes the three dimensional Heisenberg universality class.\cite{n20}

\section{Summary}

We have calculated the spin stiffness of the isotropic Heisenberg antiferromagnet on the stacked triangular geometry using
both a MC heat bath and BHM method. The spin stiffness has the advantage that it measures the rigidity of the ordered phase 
in response to a virtual twist without
having to specify the order parameter. The results obtained from both numerical approaches agree and predict a continuous phase transition which
belongs to the new chiral universality class proposed by Kawamura.

\begin{acknowledgments}
This work was supported by the Natural Sciences and Research Council of Canada and the HPC facility at the
University of Manitoba.
\end{acknowledgments}

%
%
\bibliography{p1}

\end{document}